\documentstyle[12pt,aaspp4,psfig]{article}
\newcommand{\etal}{{et al.}}
\newcommand{\halpha}{H$\alpha$}

\lefthead{Cram \etal}
\righthead{Faint radio galaxies}

\begin{document}

\title{STAR FORMATION RATES IN FAINT RADIO GALAXIES} \author{L. Cram
  \& A. Hopkins} \affil{School of Physics, University of Sydney 2006,
  Australia \\ 
  l.cram@physics.usyd.edu.au,a.hopkins@physics.usyd.edu.au} \and
\author{B. Mobasher \& M. Rowan-Robinson} \affil{Astrophysics Group,
  Imperial College, London SW7 2BZ, UK \\ b.mobasher@ic.ac.uk,
  m.rrobinson@ic.ac.uk}
\slugcomment{To be submitted to Astrophysical Journal}

\begin{abstract}
  The decimetric radio continuum luminosity of a star-forming galaxy
  appears to be directly proportional to the rate of formation of
  supernovae in the galaxy. Since decimetric radiation does not suffer
  significant extinction and is not directive, radio luminosities may
  thus provide a particularly straightforward way to determine the
  current rate of star formation. Using a sample of over 700 local
  galaxies we confirm the utility of the radio luminosity as a measure
  of star formation rate by showing concordance with the rates
  predicted by U-band, \halpha, and far infrared luminosites. We also
  show that there are systematic discrepancies between these various
  indicators, suggesting that the \halpha~luminosity may underestimate
  the star formation rate by approximately an order of magnitude when
  the star formation rate is $\gtrsim$ 20 M$_\odot$ yr$^{-1}$. We use
  this calibration and the measured radio luminosities of sub-mJy
  radio sources to infer the star formation rate in approximately 60
  star-forming galaxies at moderate ($z \gtrsim 0.1$) redshifts, both
  as the actual rate and as the fraction of the existing mass of stars
  in the galaxy. For some of these objects the inferred current rate
  of star formation could increase the stellar mass in the galaxy by
  approximately 10\% over an interval of $\approx 30$ Myr.
\end{abstract}


\section{Introduction}
In the vast majority of extragalactic radio sources with 1.4 GHz flux
density S$_{1.4} \gtrsim 5$~mJy the radio emission appears to be due
to the existence of a nuclear `monster' or `engine.'  These sources
are members of the class of powerful radio galaxies, emitting a
spectral power density $P_{1.4} \gtrsim 10^{23}$~W~Hz$^{-1}$ (e.g.
Condon \markcite{Con:89}1989, especially his Fig. 12; Wall \& Jackson
\markcite{Wal:97}1997).  Their optical hosts are intrinsically bright
elliptical galaxies with optical luminosities of the order of $5
\times 10^{37}$~W, often with red colours (e.g., Kron, Koo \&
Windhorst \markcite{Kro:85}1985) and occasionally with high-excitation
optical emission line spectra (e.g. Hine \& Longair
\markcite{Hin:79}1979).  While the nuclear monster may stimulate some
nuclear star formation, or vice versa, there is not a tight
relationship between the radio power and either the number of stars or
the current rate of star formation in the host galaxy.

Many of the remaining extragalactic radio sources with S$_{1.4}
\gtrsim 5$~mJy, as well as a large fraction of fainter sources, belong
to a different population. These have optical counterparts that are
spiral (if $V \lesssim 18.5$) or blue peculiar (if $V \gtrsim 18.5$)
galaxies with a wide range of optical luminosities, often displaying
evidence of ongoing star formation (Kron \etal~\markcite{Kro:85}1985;
Benn \etal~\markcite{Ben:93}1993).  For the more luminous members of this
population it is known that there is a tight correlation between the
far-infrared (FIR) luminosity and the 1.4 GHz radio power (reviewed by
Condon \markcite{Con:92}1992).  Astrophysical interpretations of this
correlation usually identify both the FIR and the radio emission as
the consequence of ongoing massive star formation. As a corollary, the
opportunity exists to determine the current rate of massive star
formation from measurements of the FIR or radio luminosity.

The astrophysical significance of this opportunity is heightened by
the fact that very deep radio surveys reveal large numbers of
star-forming galaxies at redshifts well beyond $z \approx 0.1$, at an
epoch where there is growing evidence that some classes of galaxies
experience star formation at a rate higher than in the local Universe.
In this paper we explore the calibration of the relationship between
radio luminosity and star formation rate using a sample of local
star-forming galaxies with measured radio flux densities, and then
apply the calibration to determine the current star formation rate in
a sample of galaxies at $0.1 \lesssim z \lesssim 0.5$.

\section{Star Formation Rates}
Several different measures have been used to estimate the current star
formation rate in galaxies, including the far-UV or U-band magnitude,
the strength of Balmer line emission, the power radiated in the FIR,
and the radio luminosity (see, e.g., Gallagher \& Hunter
\markcite{Gal:87}1987; Ferrini \markcite{Fer:97}1997). Each indicator
has been the subject of many previous investigations, but even so the
complexity of the astrophysics underlying each one leads to a
considerable degree of uncertainty in its use as an estimator of the
star formation rate.

In this section we present a simple account of the physical basis of
selected indicators and check empirically their utility using a sample
of nearby galaxies for which the radio flux density and several other
indicators are available.  We have sought to base the comparision on
similar physical assumptions, most notably by adopting similar initial
mass functions (IMF), similar models for the time-dependence of the
star formation rate (SFR), and similar models of stellar population
evolution.  The calibrations quoted below can be traced wherever
possible to an initial mass function (IMF) which varies as $\Psi(M)
\sim M^{-2.5}$ between limits of 0.1 and 100 M$_{\odot}$, and to
calculations based on the galaxy evolution models of Bruzual \&
Charlot \markcite{BrU:93}(1993) and of Fioc \& Rocca-Volmerange
\markcite{Fio:97}(1997).  Where necessary, scaling has been
applied to accommodate different physical assumptions and to produce a
homogeneous set of indicators. This scaling introduces potential
uncertainties of about a factor of 2 between the {\it ab initio}
calibration of the various indicators but could be adjusted if found
to lead to inconsistencies.

Radio continuum emission at 1.4~GHz from star-forming galaxies is
mainly synchrotron radiation produced by relativistic electrons.  It
has long been acknowledged that supernovae could play the dominant
role in accelerating these electrons (Biermann 1976; Kirk
\markcite{Kir:94}1994).  This view has been reinforced by the
discovery of the tight correlation between radio continuum and FIR
emission (reviewed by Condon \markcite{Con:92}1992), indicating that
massive stars may control both radiation mechanisms. The implication
is that the supernova rate determines the non-thermal radio
luminosity.

At first sight there is a serious problem with this interpretation,
since the total radio luminosity of a galaxy divided by the typical
luminosity of a supernova remnant implies supernova rates that are too
high (Biermann \markcite{Bir:76}1976; Gehrz, Sramek \& Weedman
\markcite{Ger:83}1983; Condon \& Yin \markcite{Con:90}1990).  The
problem can be resolved consistently with our understanding of
supernova shock acceleration mechanisms by assuming that accelerated
electrons, and perhaps the acceleration process itself, endure far
longer than the $\sim 2 \times 10^4$ yr lifetime of detectable
Galactic supernova remnants (Condon \& Yin \markcite{Con:90}1990). In
this case the relationship between the radio luminosity and the
supernova rate can be calibrated using the Galactic values of $L_{1.4}
\sim 2.3 \times 10^{22}$ W Hz$^{-1}$ and $\nu_{SN} \sim 1/43$
yr$^{-1}$.  From this Condon \markcite{Con:92}(1992) estimates the
star formation rate (SFR) of stars massive enough to form the
supernovae which contribute to radio emission (i.e., $M \ge 8$
M$_{\odot}$,) which can be adjusted using the above model for the IMF
to yield
\begin{equation}
  SFR_{1.4} (M \ge 5 M_{\odot}) = \frac{L_{1.4}}{4.0 \times
    10^{21}~{\rm W Hz}^{-1}}~M_\odot {\rm yr}^{-1}.
\end{equation}
P\'{e}rez-Olina \& Colina \markcite{Per:95}(1995) present a theory of
the relation between SFR and radio luminosity for extreme starbursts
which rests on the suggestion that the non-thermal emission from such
galaxies may be dominated by radio supernovae with lifetimes of 100 yr
or so, rather than remnants with lifetimes exceeding 20 000 yr. Their
model predicts a SFR within a factor of $\sim$2 of that given by
equation (1), since the estimates involve a direct trade-off between
the shorter duration but higher flux density of the emission from
radio supernovae when compared with supernova remnants. This
equivalence (and indeed the form of eq. [1]) requires that the star
formation rate be constant over the lifetime of the least-massive
supernova progenitors ($\approx 10^{7.5}$ yr) and the estimate is
based on a burst of star formation of this duration.

Balmer line emission from star-forming galaxies is formed in H II
regions ionised by early-type stars.  Kennicutt
(\markcite{Ken:83}1983a) has determined the theoretical relationship
between the \halpha~luminosity and the current rate of star formation
in a galaxy in a form corresponding to
\begin{equation}
  SFR_{H\alpha} (M \ge 5 M_{\odot}) = \frac{L(H\alpha)}{1.5 \times
    10^{34}~{\rm W}} M_{\odot} {\rm yr}^{-1}.
\end{equation}
An upward adjustment of 1.1 magnitude in the luminosity has been
factored in to this equation to compensate for the mean extinction in
\halpha~as suggested by Kennicutt (1983a), who also emphasises that
variations in extinction imply that {\it individual} measurements of
\halpha~luminosity must be treated with caution.  Kennicutt points out
that the statistical properties of a sample will be considerably more
accurate than measures of individual objects, a view supported by the
robustness of the data shown in Figure 1 (see below).

FIR emission from star-forming regions is due to the absorption of
stellar photospheric radiation by grains, with subsequent re-radiation
as thermal continuum in the far infrared. A simple theory relating the
FIR power of a galaxy to its current massive star formation rate can
be based on the proposition that essentially all of the UV and much of
the blue radiation from massive stars is absorbed by grains, with the
associated thermal re-radiation appearing as emission in the 40 -- 120
$\mu$m band (e.g., Xu \markcite{Xu:90}1990).  From these ideas Condon
\markcite{Con:92} (1992) derives a star formation rate equivalent to
\begin{equation}
  SFR_{FIR} (M \ge 5 M_{\odot}) = \frac{L_{60 \mu m}}{5.1 \times
    10^{23}~{\rm WHz}^{-1}} M_\odot {\rm yr}^{-1},
\end{equation}
Thronson \& Telesco (\markcite{Thr:86} 1986) derive a rate that is
$\sim 2.3$ times larger than this. The difference is due their
inclusion of the effect of dispersal of the placental star-forming
cloud, and gives a fair indication of the unavoidable uncertainty in
such theories.  As noted by Bothun, Lonsdale \& Rice \markcite{Bot:89}
(1989), there is a `lingering ambiguity' concerning the relative
importance of old disk stars and newly formed stars in heating grains
in normal disk galaxies, and several authors (e.g., Lonsdale \& Helou
\markcite{Lon:87}1987) have argued that radiation from stars less
massive than 5$_{\odot}$ will contribute significantly to FIR emission
from galactic disks. This casts doubt on the quantitative reliability of
equation (3).

The use of far-UV or U-band luminosities to infer the current star
formation rate rests on the idea that UV emission contains a
substantial contribution of light from the photospheres of young,
massive stars (e.g., Cowie \etal~\markcite{Cow:97}1997; Mobasher \&
Mazzei \markcite{Mob:97}1997).  However, the calibration of a UV
indicator is less straightforward than those listed above. Consider,
for example, the evolution of the U-band luminosity following the
sudden formation of 1 M$_{\odot}$ of stars with an IMF that varies as
$M^{-2.5}$. The spectral synthesis models of Bruzual \& Charlot (1993)
and Fioc \& Rocca-Volmerange (1997) predict that the U-band absolute
magnitude brightens to $U \approx -0.7$ after $10^{6.5}$ yr, and fades
steadily thereafter reaching $U \approx 1.7$ after $10^{7.5}$ yr.  A
starburst of $10^{7.5}$ yr would then lead to the relationship $SFR_U
(M>5M_{\odot}) \approx L_U / 1.4 \times 10^{21}$ M$_{\odot}$
yr$^{-1}$.  However, substitution of typical U-band luminosities of
disk galaxies into this relation will give values of SFR that appear
to be unreasonably high. The error is due to the large numbers of
relatively old stars that also contribute to the actual U-band
luminosity of disk galaxies. To avoid this problem, we have used the
relationship between SFR and far-UV luminosity given by Cowie
\etal~(1997), scaled from 250 nm to the U-band using relevant
synthesized spectra, to derive
\begin{equation}
  SFR_{U} (M \ge 5 M_{\odot}) = \frac{L_U}{1.5 \times 10^{22}~{\rm W
      Hz}^{-1}} M_{\odot} {\rm yr}^{-1}.
\end{equation}
The model used by Cowie \etal~assumes a steady rate of star formation
over a long interval, and is somewhat sensitive to the assumed history
of star formation (see also Gallagher \& Hunter \markcite{Gal:87}1987;
Kennicutt, Tamblyn \& Congdon \markcite{Ken:94}1994). It has the great
advantage of being directly related to one of the favored methods of
determining the SFR in high redshift ($z \gtrsim 1$) galaxies (Cowie
\etal~1997).

\placefigure{Fig1} \placetable{Table1}

An intercomparison of the rates predicted by equations (1) - (4) for a
sample of local galaxies is illustrated in Figure 1, which plots the
SFR deduced from \halpha, FIR, and U-band luminosities against the SFR
deduced from the 1.4 GHz luminosity. We have used thirteen `reference'
samples to test the relations, as summarized in Table 1 which lists
the sources of the data, the reference code used in Figure 1, and the
number of galaxies from each source for which the designated SFR
indicators have been located.

The data of Kennicutt \& Kent \markcite{KaK:83}(1983), Romanishin
\markcite{Rom:90}(1990), Kennicutt \markcite{Ken:92}(1992), Lehnert \&
Heckman \markcite{Len:96}(1996), and Young
\etal~\markcite{You:96}(1996) were chosen because these authors
tabulate a relatively large number of galaxy-integrated
\halpha~luminosities.  Eales \etal~\markcite{Eal:88}(1988) and Condon
\etal~\markcite{Con:91b}(1991) were chosen to give good coverage of the
more luminous IRAS galaxies, albeit without \halpha~ data, while Leech
\etal~\markcite{Lee:88}(1988) and Allen \etal~ \markcite{All:91}(1991)
provide coverage of a broad range of IRAS galaxies.  Gallagher, Hunter
\& Tutukov \markcite{Gal:84}(1984), Kennicutt
\etal~\markcite{Ken:87}(1987), Caldwell \etal~\markcite{Cal:91}(1991),
and Alonso-Herrero \etal~ \markcite{Alo:96}(1996) cover smaller
samples of galaxies of particular interest in studies of star
formation.

The \halpha~equivalent widths or luminosities have been taken from the
original papers. In an attempt to correct for any slit losses where
the \halpha~data have been obtained spectroscopically, we have derived
the equivalent width from the published \halpha~data and then used
published optical aperture photometry to estimate the total
\halpha~flux. Many \halpha~measurements include contamination from the
nearby [N~II] line, and where required we have multiplied the published
flux by a factor of 0.75 to account for this effect (Kennicutt
\markcite{Ken:92}1992).  The NASA Extragalactic Database (NED) has
been consulted to obtain U-band magnitudes and most of the necessary
redshifts, as well as the B-band or V-band photometry used to convert
\halpha~equivalent widths to flux densities, and the FIR flux density
when not already available in the original publication.  If necessary,
values of radio flux densities taken from the original papers have
been converted to 1.4 GHz using a spectral index of $\alpha=+0.8$
($S_\nu \propto \nu^{-\alpha}$).  The on-line NRAO/VLA Sky Survey
(NVSS) database (Condon \etal~\markcite{Con:98}1998) was used to
obtain 1.4 GHz flux densities where they were not already published.
A point is plotted in Figure 1 whenever a galaxy has a published radio
luminosity and at least one other luminosity -- we do not require that
{\it all} SFR indicators be available before plotting a galaxy.  Data
for a galaxy in the original references has been plotted only if its
1.4 GHz flux density is available: the few radio-loud objects in the
original references are also plotted in the appropriate diagrams.  All
measurements have been adjusted to correspond to a Hubble constant
$H_{0} = 75$ km s$^{-1}$ Mpc$^{-1}$.

\placefigure{Fig2}

Figure 2 plots the radio/\halpha~data for two samples of {\it distant}
star-forming galaxies, namely the objects classified as some variant
of `*' in Table 3 of Benn \etal~(1993) and the objects classified as
`Type A' in the optical follow-up of the Phoenix survey (Hopkins
\etal~\markcite{Hop:98b}1998, Hopkins \markcite{Hop:98a}1998). These
classifications, based on optical colours and the emission line
spectra, select the star-forming galaxies and reject the classical
radio galaxies for which radio luminosity is not a measure of the SFR.
There are obvious similarities between this plot and the corresponding
radio/\halpha~plot for the nearby sample.

\section{Discussion}
We consider firstly the `nearby' sample in which several different
estimators of SFR are often available for a single galaxy.  Figure 1
indicates that SFR estimates based on the various indicators are in
broad agreement with one another, but also points to the existence of
systematic differences between different estimators of the SFR. There
are also several objects in which at least one indicator is
significantly discordant.

The tightness of the relationship between the star formation rates
estimated from 1.4 GHz and 60 $\mu$m data reflects the well-studied
radio/FIR correlation.  Estimates of star-formation rates obtained
from the two frequencies agree closely over more than four orders of
magnitude. There is a tendency for objects with $SFR_{1.4} < 0.2$
M$_{\odot}$ yr$^{-1}$ to have relatively high values of $SFR_{FIR}$,
perhaps revealing some heating due to the general interstellar
radiation field of older disk stars, without associated supernovae due
to younger massive stars (see also Lonsdale \& Helou
\markcite{Lon:87}1987).  A notable exception to the correlation is the
very small number of galaxies in which the radio prediction is
unusually high. These objects include NGC 4374 (M84; 3C 272.1) from
Kennicutt \& Kent (1983), IRAS 0421+040 from Eales \etal~(1988), and
Mrk 3 (4C+70.05) from Kennicutt (1992).  In each case there was clear
prior evidence that the radio emission is not related to star
formation, and the objects have been retained in our data set only to
illustrate the effect of `radio-loud' galaxies.

The existence of these objects points to the potential problem that
1.4~GHz luminosities may be `contaminated' by radio emission arising
from a nuclear monster, rather than star formation. In deducing the
SFR from radio flux densities this contamination can be minimized by
applying the following principles: (1) optical colours and optical
spectral line ratios sucessfully delineate the class of objects with
emission dominated by star formation (e.g. Benn \etal~1993), (2) the
fraction of galaxies which owe their radio emission to star formation
rises quickly with decreasing flux density in the sub-mJy radio source
population (e.g. Kron \etal~1985; Wall \& Jackson 1997), (3) any
optical emission lines in classical radio galaxies with relatively low
radio powers are relatively weak but lie in the high-excitation part
of the relevant diagnostic diagram (e.g. Baum \& Heckman
\markcite{Bau:89}1989), and (4) the bi-modal character of the
radio-loud/radio-quiet dichotomy allows the use of the radio/optical
flux density ratio to exclude the most powerful radio-loud objects
(e.g, Sopp \& Alexander \markcite{Sop:91}1991, their Fig 1.)

Figure 1 shows that estimates of the SFR based on the widely-used
\halpha~luminosity indicator tend to be more scattered (relative to
$SFR_{1.4}$) than the $SFR_{1.4}/SFR_{FIR}$ correlation. They also
tend to lie about a factor of 10 above the trend defined by the 1.4
GHz/60 $\mu$m estimates when SFR $\lesssim$ 0.1 M$_\odot$ yr$^{-1}$,
and about a factor of 10 below the trend when SFR $\gtrsim$ 20
M$_\odot$ yr$^{-1}$.  The systematic deviations, which could have
important implications for studies of the galaxy-to-galaxy variation
of star formation rates, can be seen also in Figure 1 of Devereaux \&
Young \markcite{Dev:90}(1990) and Figure 8 of Young \etal~(1996),
although these authors did not emphasise the discrepancy.  On the
other hand, investigations spanning a smaller range of luminosities
(e.g. Kennicutt \markcite{Ken:83}1983b; Lonsdale \& Helou 1987) have
sometimes infered that the relationship is close to strict
proportionality.  Devereaux \& Young (1990) argue that the general
correlation between \halpha~and FIR results constitutes `strong
evidence' for the view that the \halpha~and the FIR luminosity are
produced mainly by massive stars in galaxies with high FIR luminosity.
Lonsdale \& Helou (1987) argue on the other hand that a parallel
correlation between the FIR and B-band luminosity implies that the FIR
emission in most spiral galaxies is `dominated' by emission due to
cool dust which is heated in part by low-mass stars. Even in this
case, however, we would expect the FIR luminosity to be increasingly
dominated by ionising radiation in the presence of vigorous star
formation.

The systematic over-estimate of the SFR derived from \halpha~relative
to the 1.4~GHz estimates at {\it low} values of SFR is related in part
to the difficulty of determining the zero-point level for
\halpha~emission from H II regions in the presence of stellar
\halpha~emission or absorption. Furthermore, since the radio/FIR
correlation can be `linearized' by allowing for the escape of cosmic
rays from galaxies with weak star formation (Condon, Anderson \& Helou
\markcite{Con:91a}1991), we must also entertain the possiblity that the
SFR estimate based on $L_1.4$ is actually too low in these objects.

The deviation at high radio luminosities, as well as the increasing
scatter in the correlation, could result from a relatively large
amount of extinction in those objects undergoing the most vigorous
star formation, or from the absorption by of Lyman photons by dust, or
from an IMF which weights differently the high-mass stars mainly
responsible for \halpha~and the lower mass stars which dominate the
supernova numbers (or to a combination of these factors). The fact
that the estimates based on the edge-on sample of Lehnert and Heckmann
(1996 - the open triangles in Figure 1) fall systematically low
supports the view that extinction plays a particularly significant
role. It is notable that the value of SFR derived from \halpha~tends
to be bounded above by a rate of $\approx 25$ M$_{\odot}$ yr$^{-1}$,
corresponding to a luminosity of about $3 \times 10^{35}$ W Hz$^{-1}$.

Estimates of the SFR based on U-band observations exhibit considerable
scatter with respect to the radio estimates. Like the estimates based
on \halpha~they tend to be higher at low SFR and lower at high SFR. As
noted above, even old stellar populations emit U-band light (e.g.,
Bruzual \& Charlot \markcite{Bru:93}1993, their Fig 1) providing a
possible explanation of the trend observed at low values of SFR.  The
trend for relatively low SFR values towards the right of the figure
could reflect enhanced extinction in vigorous starbursts. As with
\halpha~there appears to be an upper limit to the SFR derived from
U-band measurements.  The discrepancies highlighted here need to be
further investigated, especially given the increasing use of
broad-band far-UV luminosities to infer star formation rates in
galaxies with very high redshifts (z $\gtrsim$ 1; e.g.  Cowie
\etal~\markcite{Cow:97}1997).

The points shown in Figure 2 correspond to galaxies selected by very
deep surveys in the radio continuum. Of the star-formation indicators
discussed above, only radio and \halpha~luminosities are available at
present for these samples.  Moreover, the \halpha~values have been
derived from noisy spectra which were obtained to determine redshifts
rather than line fluxes, and hence they are expected to show
significant scatter. The expedient of rejecting radio-selected objects
that are red and/or show absorption line spectra has led to a sample
which follows a trend that is essentially indistinguishable from the
`local' sample. It is thus reasonable to conclude that (1) the
`shortfall' in the \halpha~estimates of the SFR for the distant sample
is similar to that in the local population and is likely to arise for
the same reasons, and (2) the radio luminosities of these types of
galaxies provide estimates of their current star formation rates
according to equation (1). A corollary of this is that several faint
radio-selected galaxies have star formation rates comparable with
those of the intrinsically luminous IRAS galaxies.  There is therefore
considerable promise that deep radio surveys will reveal significant
numbers of objects similar to the luminous IRAS galaxies, but lying at
higher redshifts.

While the SFR of a galaxy is of interest in its own right, the ratio
of the current SFR to the total number of stars that have been formed
in a galaxy offers additional insight into the potency of the current
rate of star formation.  We can estimate the total mass of stars
formed in the faint radio galaxies, $M_{TOT}$, by using their $R$-band
luminosities in the relation
\begin{equation}
  M_{TOT} = \frac{L_R}{3.4 \times 10^{11}~{\rm W Hz}^{-1}} {\rm
    M_{\odot}}.
\end{equation}
Here, we have used the approximation (deduced from Bruzual \& Charlot
1993) that steady star formation, with a standard IMF and occurring
over a period of more than 1 Gyr, will produce an $R$-band
mass-to-light ratio of $R \approx 5$~mag~M$_{\odot}^{-1}$.  Since this
approach fails to account for the contribution of any {\it current}
starburst to the $R$-band luminosity it will underestimate the
fractional amplitude of very strong bursts.

\placefigure{Fig3}

Figure 3 exhibits, for the sample shown in Figure 2, the relationship
between the SFR (of massive stars) deduced from the radio luminosity
and the total mass of stars deduced from equation (5). Lines are drawn
to illustrate the locii of points for which the characteristic time
$\tau = M_{TOT} / SFR_{1.4}$ is constant. There is a tendency for
galaxies that have already formed many stars to support a higher rate
of current star formation. There is also a wide scatter in the ratio
of SFR to total mass at any chosen size, not all of which is due to
errors of observation. There are about a dozen galaxies with a stellar
mass $M \approx 10^{10}$ M$_{\odot}$ and a star formation rate $SFR(M
> 5 M_{\odot}) \geq 10$ M$_{\odot}$ yr$^{-1}$. For such galaxies, the
current burst of star formation is likely to increase the stellar mass
by more than 10\% if sustained for a time exceeding $10^7$~yr,
implying an event of considerable significance in the development of
the system. These objects are reminiscent of the IRAS galaxies that
have high values of $L_{FIR}/L_B$ (e.g., Sanders \& Mirabel
\markcite{San:96}1996, their Section 2.2).

\section{Prospects}
The utility of decimetric radio luminosity as a measure of the star
formation rate in a galaxy relies on the hypothesis that the radio
luminosity is directly proportional to the supernova rate.  The
radio-FIR correlation can be cited as support for this hypothesis,
although the `shortfall' of \halpha~ and U-band predictions (Figure 1)
at high radio luminosities might be seen as evidence against it.
Given the potential rewards of applications of the relation, the
astrophysical interpretation of the phenomena controlling the form of
Figure 1 must be explored further.  There appears to be particular
concern regarding the interpretation of values of SFR inferred from UV
luminosities. It is evident that an exploration of correlations
between {\it deviations} from the trend lines shown in Figure 1,
combined with additional two-color optical and FIR data, could
help to elucidate these phenomena.

Armed with the capacity to determine star formation rates from radio
luminosities, we are in a position to probe the current star formation
rates of galaxies to redshifts well beyond $z = 0.1$, provided that we
can obtain optical identifications and thence redshifts. Optical
photometric and spectroscopic observations also help to confirm that
the galaxy is not host to a `monster' and to eliminate the possibility
this it is radio loud.  {\it Radio selection} of the candidates will
preferentially reveal objects with high rates of current star
formation.

We are presently extending the radio frequency sensitivity of the
Phoenix survey (Hopkins \etal~\markcite{Hop:8b}1998) using the
Australia Telescope, and seeking to obtain a large number of redshifts
(approaching 500) for the identified radio sources using the 2dF fiber
spectrograph on the Anglo-Australian Telescope. These data will
significantly enhance our understanding of star formation in the
region $0.1 \lesssim z \lesssim 1$.

\acknowledgements This research has used the NASA/IPAC Extragalactic
Database (NED) which is operated by the Jet Propulsion Laboratory,
California Institute of Technology, under contract with the National
Aeronautics and Space Administration. The research has also used the
NRAO VLA Sky Survey (NVSS) Database.  The National Radio Astronomy
Observatory is a facility of the National Science Foundation operated
under cooperative agreement by Associated Universities, Inc. The
research has been supported by the Australian Research Council and the
Science Foundation for Physics within the University of Sydney.

\clearpage

\figcaption[FIG1.eps]{Comparison of star formation rates deduced from
  1.4 GHz luminosities (horizontal axis) with the rates deduced using
  the other luminosities indicated. Symbols are coded by the initials
  of the author's last names and the year of publication, and
  cross-referenced in Table 1. \label{FIG1}}

\figcaption[FIG2.eps]{Comparison of star formation rates deduced from
  1.4 GHz luminosities (horizontal axis) with the rates deduced from
  measured \halpha~luminosities, for two samples of faint radio
  galaxies. Triangles denote the sample of Benn \etal~(1993) and stars
  denote the sample of Hopkins (1998) and Hopkins \etal~(1998).
  \label{FIG2}}

\figcaption[FIG3.eps]{The current star formation rate inferred from
  the 1.4 GHz luminosity versus the total mass of stars inferred from
  the R-band luminosity for the objects contained in the
  radio-selected samples of Benn \etal~(1993) and Hopkins
  \etal~(1998). From left to right, the lines indicate the locii of
  points with characteristic times $\tau = M_{TOT} / SFR_{1.4} = 10^8,
  10^9 and 10^{10}$ yr, respectively. \label{FIG3}}

\clearpage

\begin{deluxetable}{lllcccc}
  \scriptsize \tablecolumns{7} \tablewidth{0pt} \tablecaption{Sources
    of data} \tablehead { & & & \multicolumn{4}{c}{Numbers of
      Galaxies} \\ \cline{4-7} \\ \colhead{Authors} & \colhead{Code} &
    \colhead{Selection Criteria} & \colhead{1.4 GHz} &
    \colhead{\halpha} & \colhead{60 $\mu$m} & \colhead{U-band} }
  \startdata Kennicutt \& Kent~(1983) & KK83 & Opt. mag. limited,
  mostly spirals & 110 & 91 & 104 & 85 \\ Gallagher \etal~(1984) &
  GHT84 & Irr/Spirals with high SFR & 8 & 9 & 8 & 8 \\ Kennicut
  \etal~(1987) & KKHHR87 & Interacting spirals & 37 & 29 & 23 & 24 \\ 
  Eales \etal~(1988) & EWB88 & IRAS galaxies & 59 & & 59 & \\ Leech
  \etal~(1988) & LLRRWP88 & Northern hemisphere IRAS galaxies & 133 &
  65 & 131 & 2 \\ Romanishin (1990) & R90 & Spirals & 42 & 42 & 39 &
  20 \\ Condon \etal~(1991) & CHYT91 & Ultra-luminous IRAS galaxies &
  32 & & 32 & 4 \\ Allen \etal~(1991) & ANMR91 & Southern hemisphere
  IRAS galaxies & 179 & 179 & 179 & \\ Caldwell \etal~(1991) & CKPS91
  & Sa galaxies & 1 & 1 & 1 & 1 \\ Kennicutt (1992) & K92 & Nearby,
  diverse types & 55 & 50 & 47 & 40 \\ Lehnert \& Heckman (1996) &
  LH96 & IR-selected starbursts (edge on) & 28 & 13 & 27 & 19 \\ 
  Alonso \etal~(1996) & AAZR96 & \halpha-selected starbursts & 2 & 2 &
  1 & \\ Young \etal~(1996) & YKALR96 & Galaxies with CO data & 74 &
  74 & 70 & 57 \\ \\ Benn \etal~(1993) & & VLA \& WRST 1.4 GHz, deep
  surveys & 39 & 39 & & \\ Hopkins \etal~(1998) & & ATCA 1.4 GHz, deep
  survey & 24 & 24 & & \\ \enddata
\end{deluxetable}



\clearpage
\begin{figure}
\figurenum{1}
\epsscale{0.7}
\plotone{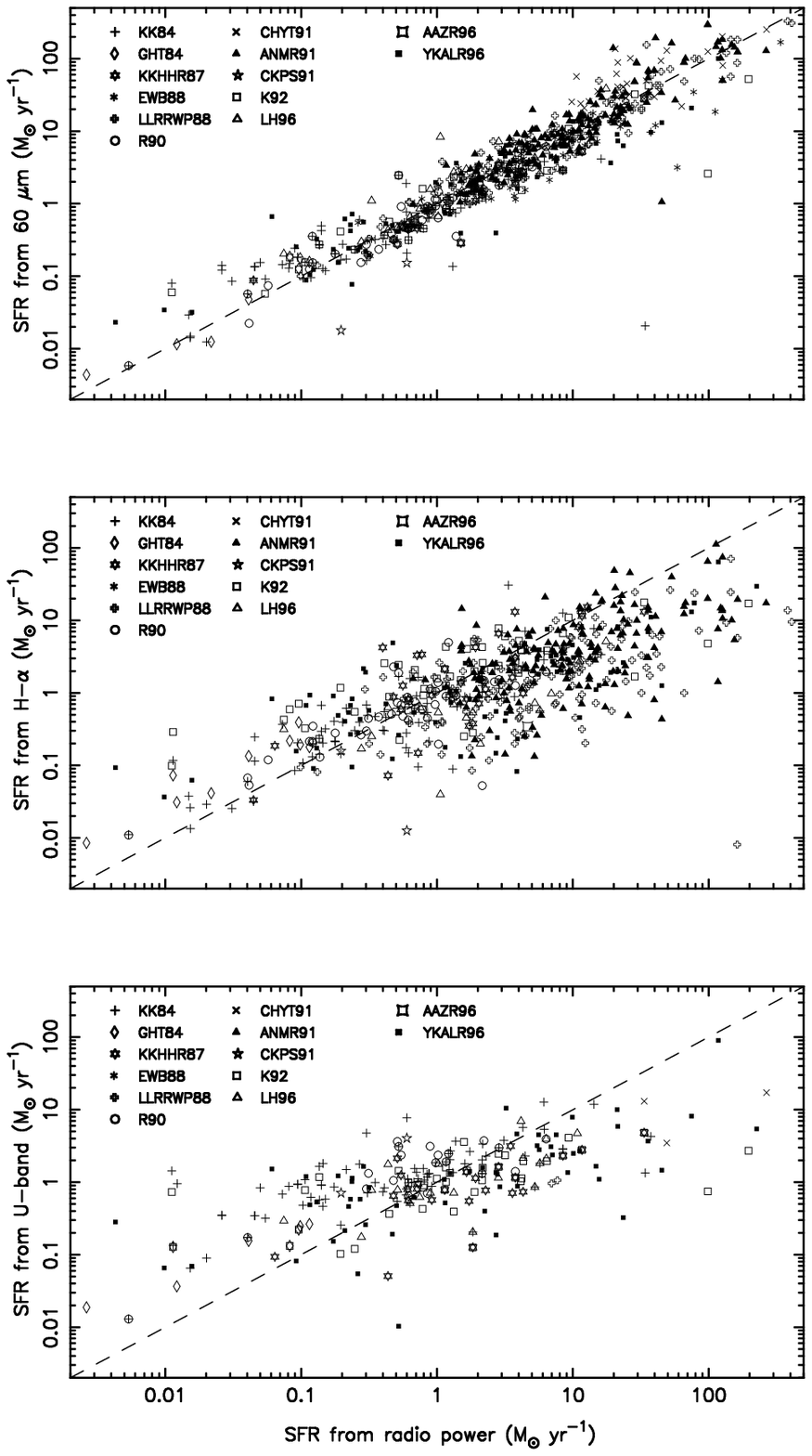}
\end{figure}

\clearpage
\begin{figure}
\figurenum{2}
\epsscale{0.7}
\plotone{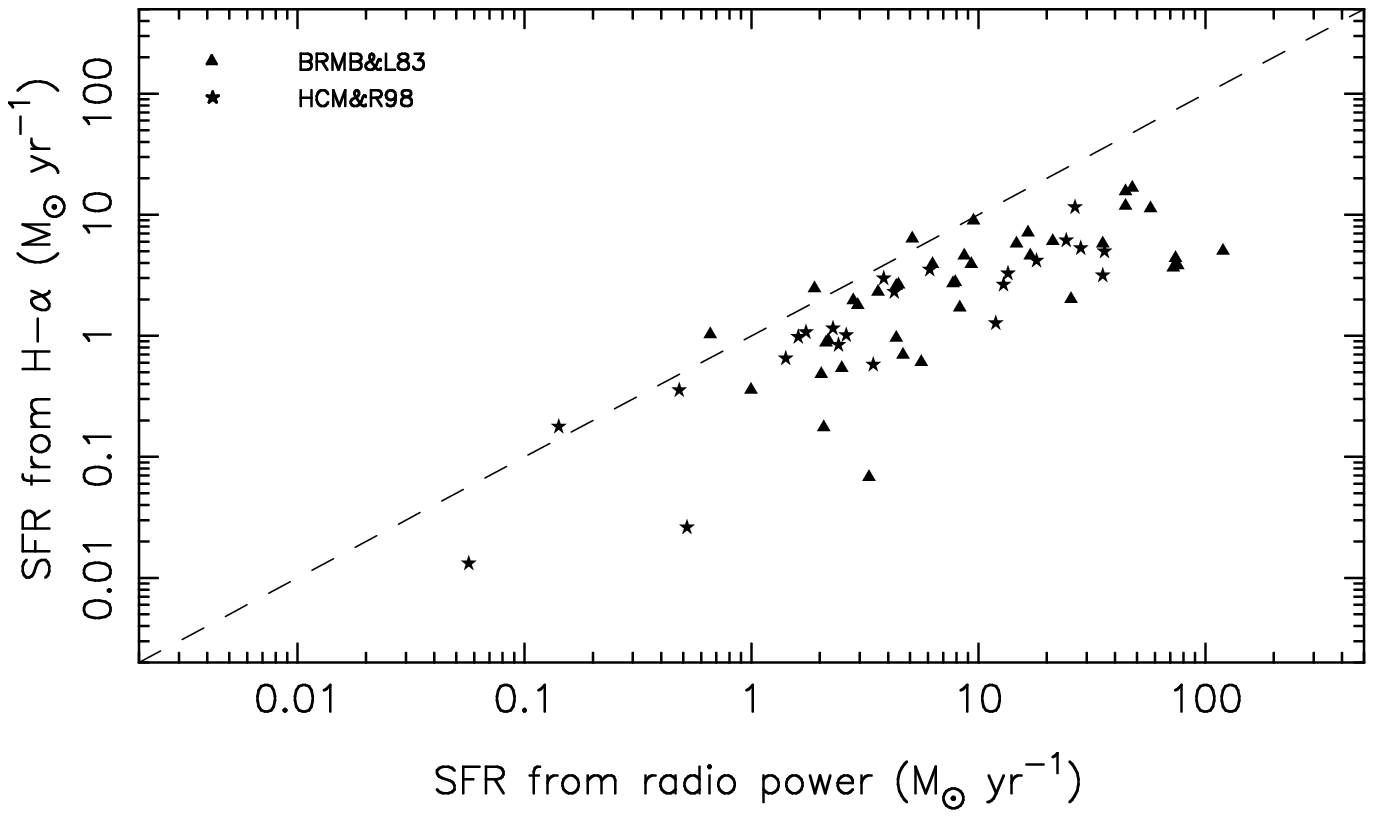}
\end{figure}

\clearpage
\begin{figure}
\figurenum{3}
\epsscale{0.7}
\plotone{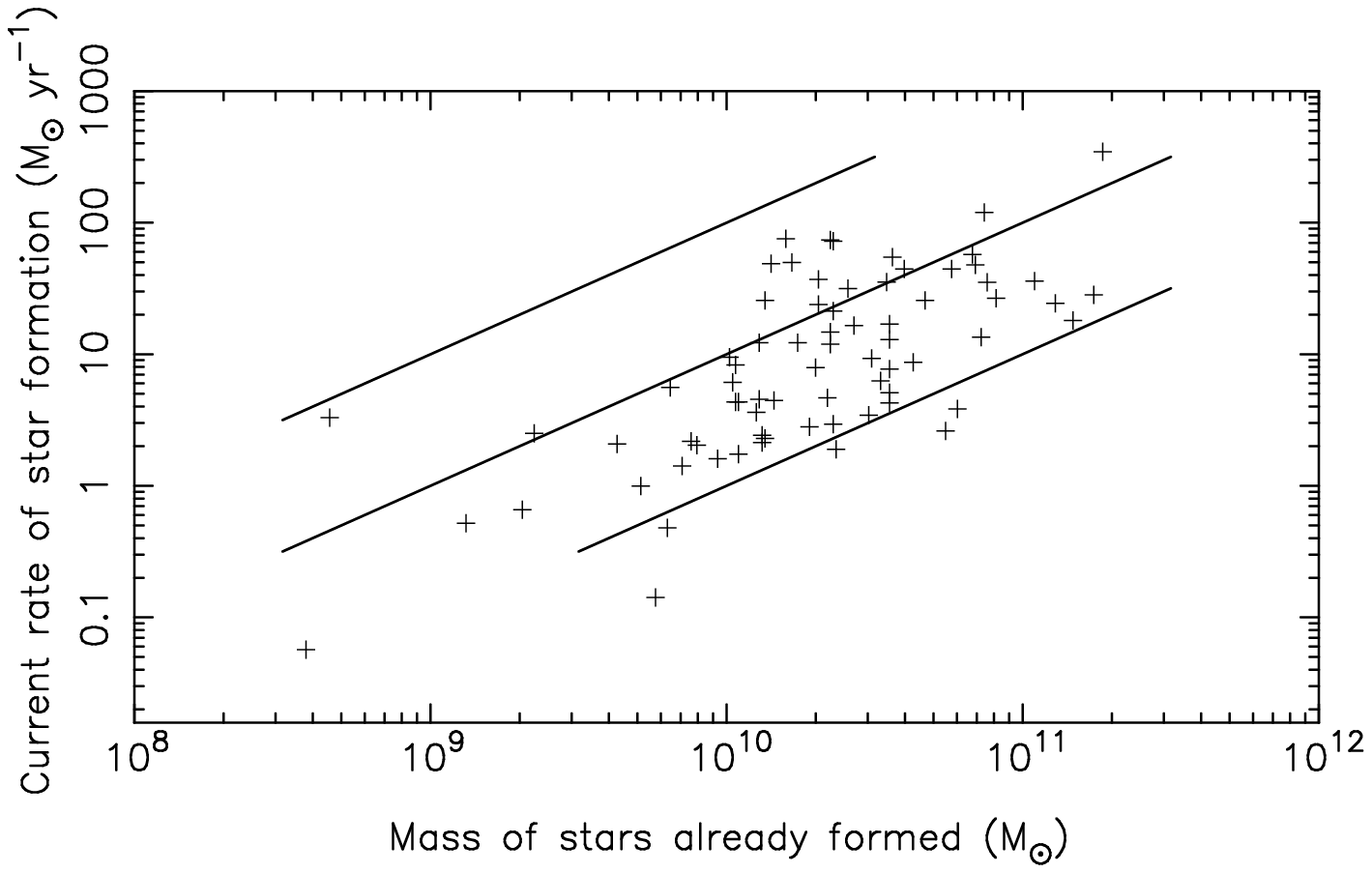}
\end{figure}

\end{document}